\documentclass[AMA,STIX1COL]{WileyNJD-v2}
\usepackage{comment,bm}
\articletype{Research Article}%

\received{}
\revised{}
\accepted{}

\begin{document}

\title{Spatial Temporal Aggregated Predictors to Examine Built Environment Health Effects}

\author[1]{Adam Peterson}

\author[2]{Jana A. Hirsch}

\author[2]{Brisa N. S\'anchez*}

\authormark{Peterson \textsc{et al}}

\address[1]{\orgdiv{Dept. of Biostatistics}, \orgname{University of Michigan}, \orgaddress{\state{MI}, \country{United States}}}

\address[2]{\orgdiv{Dept. Biostatistics and Epidemiology}, \orgname{Drexel University}, \orgaddress{\state{Pennsylvania}, \country{United States}}}

\corres{*\email{bns48@drexel.edu}}

\abstract[Summary]{
We propose the spatial-temporal aggregated predictor (STAP) modeling framework to address measurement and estimation issues that arise when assessing the relationship between built environment features (BEF) and health outcomes. Many BEFs can be mapped as point locations and thus traditional exposure metrics are based on the number of features within a pre-specified spatial unit. The size of the spatial unit--or spatial scale--that is most appropriate for a particular health outcome is unknown and its choice inextricably impacts the estimated health effect. A related issue is the lack of knowledge of the temporal scale--or the length of exposure time that is necessary for the BEF to render its full effect on the health outcome.   The proposed STAP model enables investigators to estimate both the spatial and temporal scales for a given BEF in a data-driven fashion, thereby providing a flexible solution for measuring the relationship between outcomes and spatial proximity to point-referenced exposures. Simulation studies verify the validity of our method for estimating the scales as well as the association between availability of BEFs' and health outcomes. We apply this method to estimate the spatial-temporal association between supermarkets and BMI using data from the Multi-Ethnic Atherosclerosis Study, demonstrating the method's applicability in cohort studies.}


\maketitle

\footnotetext{\textbf{Abbreviations:} BEF, Built Environment Feature; STAP, Spatial Temporal Aggregated Predictors; BMI, Body Mass Index;}

\section{Introduction}
\label{sec1}

 An expanding body of research is focused on quantifying how exposure to built environment characteristics impact health \citep[e.g.,][]{booth2005obesity,charreire2010measuring,schipperijn2015longitudinal,davis2009proximity,roux2016impact,kaiser2016neighborhood}. The built environment refers to the human-made space in which humans live, work and recreate on a day-to-day basis \citep{roof2008public}. As such, features of these environments constrain and/or enable everyday choices that may contribute to the development of disease.  The features of the built environment are many - ranging from sidewalk availability and street connectivity to the geographic density and distribution of certain amenities such as community centers or businesses that can be mapped as point locations according to their address. In this paper we are primarily concerned with the latter and for simplicity refer to them as built environment features (BEFs). In the study that motivated this manuscript, for example, a review of different measures of BEFs used to characterize neighborhood environments was conducted, identifying that both physical and social environments are related to risk factors for cardiovascular disease \citep{roux2016impact}. In particular there is growing evidence to support a causal relationship between body mass index (BMI) and healthy food availability, as measured   by the number of supermarkets, fruit-and-vegetable stores, and recreational facilities in a one mile radius around participants residential locations \citep{barrientos2017neighborhood}.
 
 \par A key limitation in the current literature focused  on estimating these effects is the unknown spatial unit within which BEFs should be measured and the functional form of the relationship between BEF effects across space. 
Given this lack of knowledge, studies differ in the unit used to define neighborhood environment metrics: Circular areas (``buffers'') of varying radii, census tracts, block groups, and counties are all examples of different areal aggregations within which BEFs have been measured previously in the literature. 
Not only do these approaches make it challenging to compare and synthesize results \citep{papas2007built,chaix2005comparison,leal2011associations}, but they can also induce severe bias in the association of interest when the incorrect spatial unit is used \citep{baek2016distributed}. The latter issue has been long been recognized as the ``modifiable area unit problem''  \citep{spielman2009spatial,fotheringham1991modifiable,openshaw1996developing,james2014effects,guo2004modifiable}. An increasingly common approach is to use an individual-centered measure-- for instance the availability of BEFs within a 1 mile radius of subjects' residential address, a distance equivalent to about a 20 minute walk. This approach injects subject matter expertise about the behavior of subjects (walking speed) and contextual knowledge (walking for transportation is common) to define the area for measuring BEF availability \citep{an2012school,howard2011proximity}. However, given their \textit{a priori} nature, these approaches fail to empirically identify the most relevant spatial unit for measurement \citep{spielman2009spatial}.

\par Recent methodological work in this area has thus focused on furthering understanding of the distance at which these effects occur---also referred to as the spatial scale--in a data driven fashion. These scales are of relevance for decision making for both urban design and policy making, since they describe how the geographic distribution of specific amenities may support specific health outcomes.   Improving upon the buffer-based approach, \citep{baek2016distributed} adapted distributed lag models (DLMs)  to avoid pre-specification of the spatial scale.  Instead of buffers, DLMs use counts of BEFs within a discrete series of concentric, ring-shaped areas as predictors in a regression model. Since each count is tied to the radii of the ring shaped area, the DLM thus estimates how effects of BEFs change as a function of the distance between BEFs and study participants. However, this method is limited in that it requires  discretization of distance to form the ring-shaped areas, and does not enforce any substantively driven functional constraints - e.g. monotonicity of the effect with respect to distance between subjects and BEF locations. The DLM also cannot estimate spatial-temporal effects without a dramatic increase in the number of estimated parameters.

A related, although even less discussed, methodological issue is how BEF effects can vary with duration of exposure--what we will refer to as the temporal scale. That is, a longer exposure duration may be needed for a given BEF to have maximum impact. Some BEFs may have a larger temporal scale, depending on the health outcome and the pathway thorough which it confers its effect. For biological processes that take time to change, e.g., BMI, the units of the temporal scale may be in the order of months to years. On the other hand, physical activity may have a smaller time scale, potentially in the order of days or weeks, because even though it may take time to build a habit of walking for transportation, the decision to walk to a newly open gym can be instantaneous.  
Although there has been work on models to incorporate exposure histories\citep{bandeen1999modelling,bandeen2010cumulative}, we are unaware of any research that  systematically investigates the estimation of temporal scales in the built environment literature. 

\par This work introduces the Spatial Temporal Aggregated Predictor (STAP) model, which addresses questions about both the spatial and temporal scale at which BEFs contribute to health outcomes. It then demonstrates the utility of STAP through analysis of simulated and patient-cohort data.   Section 2 introduces the Multi-Ethnic Study of Atherosclerosis (MESA) patient cohort which represents the subject-level data used in our motivating example \cite{roux2016impact}.
Section 2 also presents the BEF point pattern data structure used by STAP and pertinent substantive choices that ground STAP modeling decisions.
In Section 3 we provide a formulation of generalized linear mixed models and how the STAP framework extends this model family. 
Section 4 discusses how the STAP model is estimated in a Bayesian paradigm and how to consider prior choices and model selection. 
Section 5 demonstrates the model in the context of simulated data, while Section 6 showcases STAP in a real world setting via analysis of MESA data.
We conclude with a discussion of future work including possible extensions to the current model in Section 7.

\section{The Multi-Ethnic Study of Atherosclerosis and Healthy food store availability}\label{sec2}
\subsection{The Multi-Ethnic Study of Atherosclerosis}
The Multi-Ethnic Study of Atherosclerosis is a longitudinal cohort of men and women from six communities in the United States that seeks to understand the determinants of cardiovascular disease (CVD) prevalence, incidence and progression \citep{bild2002multi}. 
The MESA Neighborhoods Ancillary Study  measured a rich set of time-varying characteristics of the participants' neighborhood environments to examine their influence on CVD, for the first five visits of MESA, spanning 2000-2010 \citep{roux2016impact}. These data contain participants' geocoded residential locations, Census based-characteristics, as well as the locations of a large set of amenities near study participants, among others. We focus on exposure to food stores that are considered to be supportive of a healthy diet (healthy food stores, HFS). These stores include supermarkets, which carry a large variety of produce, among other items, as well as specialty stores including produce and fish markets. Further, we focus on the North Carolina site of MESA, given that its urban context is distinct from other sites and the dominant transportation mode (car) makes it so that the spatial scale may be larger than conventionally assumed spatial scales (e.g., 1 mile). 
The health outcome of interest in our analysis is body mass index (BMI), a salient risk factor for CVD.

\subsection{Overview of Data Construction}
The analysis of point-referenced BEF data begins with the collection and classification  of the businesses and amenities near subjects into classes appropriate for the scientific question of interest. In our analysis of MESA data, we used the National Establishments Time Series (NETS) database, a commercial database containing the locations, names and descriptions of business establishments nationwide, to identify HFS and their locations.  A variety of methods have been proposed to ensure that business classifications are consistent and appropriate for the goods and services they offer \citep{auchincloss2012improving, hoehner2010concordance}. In our analysis we used an approach that combined the standard industry code, trade name, sales volume and more, to identify supermarkets \citep{kaufman2015measuring,hirsch2020business}.
\par In addition to this classification, the mapping of each BEF's address to their latitude and longitude allows for the calculation of pairwise distances between BEFs of interest and study subjects' locations. Pairwise distances can then be used to construct exposure measures. The subjects' work, residential, and other addresses may be used according to the scientific question of interest; in MESA we focused on residential addresses as our interest is in examining the residential neighborhood environment.  Similarly, while the classical Euclidean (``as the crow flies'') distance may be used, we used the street network distance as it is believed to provide a more accurate proxy of spatial access.

\par When feasible, the data can also include the time that the subject has lived or worked at the corresponding distance from the BEF, to enable the estimation of the temporal scale. In MESA, we calculated this as the difference in time, in years, between the study visit date and the date the business opened, or between the study visit date and the time the subject moved to the current address, whichever is shortest.  We use this information to estimate the temporal scale for the relationship between HFS exposure and BMI.

\par The end result of the business classification and pairwise distance calculation is a ``long'' data frame, with multiple rows per participant, each containing the distance from each subject's residential address to each business, along with the time of exposure to that particular business. In the case of a longitudinal study with multiple visits per subject, the data would consist of multiple sets of  pairwise distances and times associated with each subject's study visit (Supplementary Table 1).
\par For practical reasons, some limit will likely have to be placed on the number of businesses to include in model fitting and/or an upper limit on the distances between subject-BEF.  For instance, in large, dense cities, a participant may easily accrue more than one hundred coffee shops within a five mile radius, and businesses beyond that could be excluded.  This limit on the number of establishments or outermost distance, however, should be chosen on the basis of substantive reasoning, with an emphasis towards being more conservative - including more businesses - since this will have a greater chance of ensuring that subjects' exposure is estimated accurately. With these ideas in mind, we included HFS within 10km of the subjects' residential address in our analysis.  We view 10 km as an appropriate and  conservative inclusion distance, as a similar cutoff was used in a similar analysis \citep{baek2016hierarchical}. Given the availability of data on when businesses opened and closed, which is part of the NETS database, and because we additionally have data on MESA participants' residential history (i.e., the dates when they move to a new address), we also calculated the time in years spent exposed to these nearby HFS. This time ranged from effectively 0 to twenty years at the last MESA visit.

\subsection{Descriptive analysis of distance data in MESA}

\begin{figure}[H]
    \centering
    \caption{(a) Distribution of distances between healthy foods stores and residential locations for a sample of 50 subjects at the baseline visit (dot=median distance for each subject, lines span 2.5\% and 97.5\% percentiles), sorted by median distances. (b) Distribution of the number of HFS within network buffers of varying size; line types indicate different visits. (c) Within-subject differences in the BEF count within the buffers of varying size, comparing the exposure count at a given visit from the subject's average count.}
    \label{fig:ddescriptive}
    \includegraphics[width =.9\textwidth]{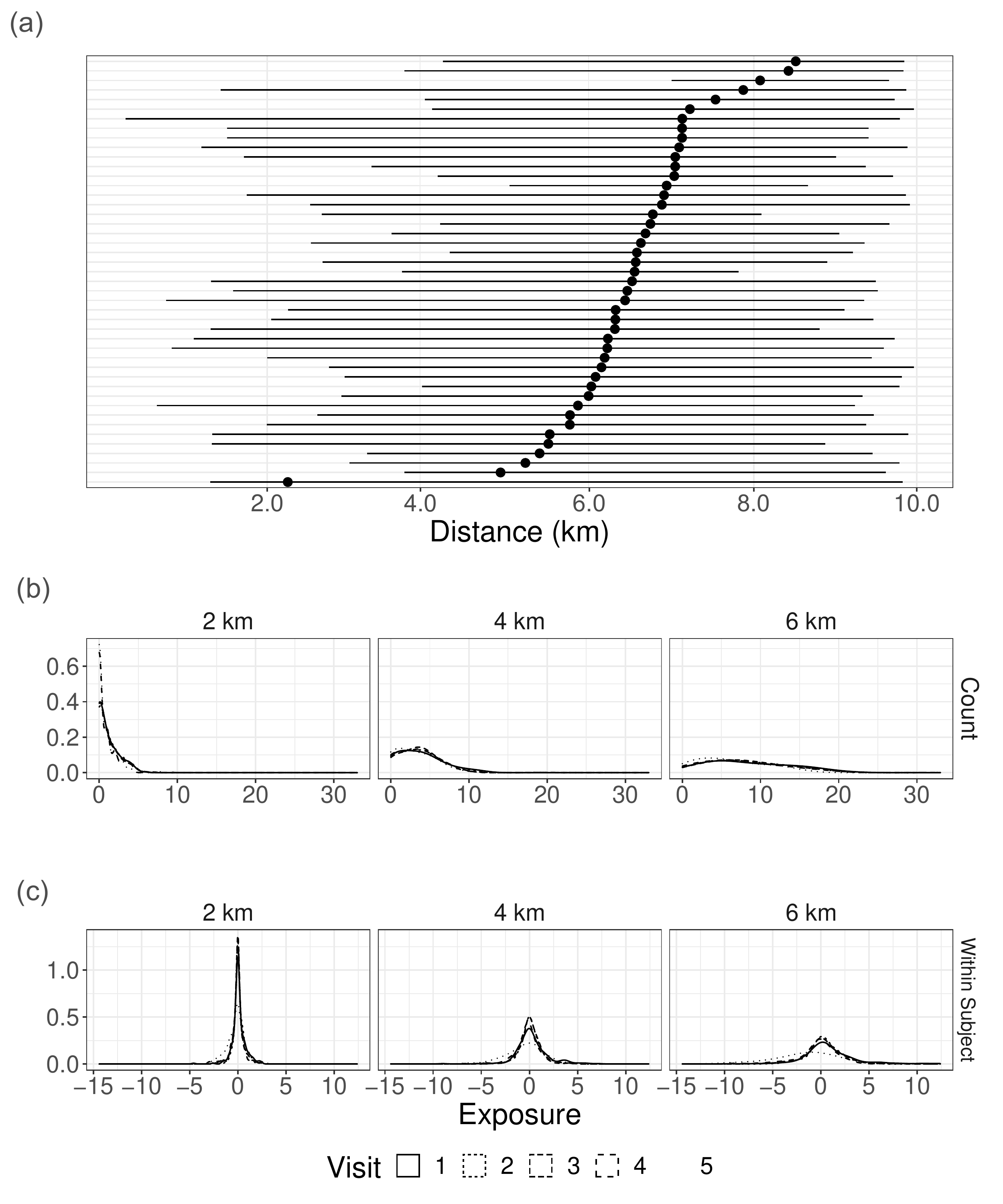}
\end{figure}

Figure \ref{fig:ddescriptive}a shows the distribution of network distances between subjects and nearby HFS for a sample of 50 subjects at their baseline visit. For example, the first subject (first from the bottom of  Figure 1a) had HFS as close as 1km, and as far as nearly 10km, but half of their HFS were within 2.5 km. This plot shows the distances that are utilized in our modeling approach described in the next section, and is informative in showing where the majority of HFS are located with respect to subjects' residence more broadly. While the first HFS is typically within 2.5 km of where subjects live (the minimum distance), the greater majority of HFS are more than 5 km away from subjects.

To further describe the data, we also use a typical exposure metric to calculate subjects' HFS exposure, namely the count of HFS within buffers of pre-determined sizes around each subject for each visit (Figure \ref{fig:ddescriptive}b). These counts are a classic exposure assessment approach, and also enable us to calculate change in exposure within subject in an intuitive fashion.  We calculate change within subject by subtracting their average HFS count across all visits from the subject's HFS exposure count at a specific visit. Figure \ref{fig:ddescriptive}c shows density estimates for the within subject change in exposure for varying buffers sizes at each visit, to give a sense of how exposure can vary across buffer sizes and visit. While there is great variability in the exposure counts across buffer size, there is relatively less variability across visits within subject. This is somewhat unsurprising given that the study period spans approximately 10 years with visits approximately 2 years apart, and, although we focused on participants who moved residences during the study period, it is likely that participants moved to neighborhoods with similar characteristics.

\section{The STAP Model}
The STAP model extends the standard generalized linear mixed model (GLMM) regression by incorporating the estimation of the spatial and temporal scale at which one or more built environment exposure affect an outcome, as well as their corresponding  coefficients.

\subsection{The Univariate Model}
\par For simplicity, we describe the model focusing on one type of BEF, HFS for example, and the mean of health outcome $Y_i$, $\mu_i \in \mathbb{R}^1$, discussing how it can be expanded later. In order to estimate the BEF effect and spatial scale, the STAP model requires pairwise distances $d$ between each subject and each BEF, in addition to a spatial exposure function,  $\mathcal{K}_s$, defined below.  
\par Let $\mathcal{D}_{i}$ be the set of aforementioned pairwise distances, $d \in \mathbb{R}^+$, between all the BEF locations and subject $i$. 
Denoting $g(\cdot),\alpha,\bm{\delta},\bm{Z}_i$ as link function, intercept, fixed effects parameters, and $i$th subject's covariate vector respectively, the model is then:
    
\begin{align*}
    g(\mu_i) &= \alpha + \beta X_{i}(\theta) + \bm{Z}_i^{T}\bm{\delta},  \tag{1}\\
    X_{i}(\theta) &= \sum_{d \in \mathcal{D}_{i}} \mathcal{K}_s \big( d,\theta^s \big ).
\end{align*}
The construction of the spatial covariate, $X_{i}$, has several implications for the interpretation of the model and permits estimation of the spatial scale that is of particular interest to built environment researchers.  
\par First, the $X_{i}$ represents subject $i$'s  cumulative exposure to the particular type of BEF, accumulated according to the spatial parameter $\theta^{s} \in \mathbb{R}^+$ and spatial exposure function $\mathcal{K}_s$.   Choosing  $\mathcal{K}_s$ to be equal to 1 up until a pre-defined distance of say, 2 kilometers, would be equivalent to counting BEFs within a pre-specified buffer of radius 2 kilometers. However, we instead let $\mathcal{K}_s$ be a non-increasing function where $K_s(0,\theta^s) = 1$ and $ \lim_{d \to \infty} \mathcal{K}_s(d,\theta^s) = 0$. This corresponds to the more realistic substantive belief that a given BEFs' maximum impact is made when a subject is as close as possible to it and that the impact decreases with longer distance. To that end, we utilize the survival functions of positive continuous random variables or the complementary error function, as they satisfy all the aforementioned constraints for $\mathcal{K}_s$. While some survival functions, such as that of the Weibull distribution, may require more than one spatial parameter, i.e., $\bm{\theta}^s = (\theta^s_1,\theta^s_2)$, for now we limit our discussion to the univariate case as the extension is straightforward and demonstrated in later sections.

Second, the coefficient $\beta$ is the estimated difference in $g(\mu_i)$ associated with one unit higher cumulative exposure. However, the construction of $X_{i}$ allows for that ``one unit'' higher to be achieved by  placing one new BEF at a distance $0$ from the subject, all else equal. Hence $\beta$ also represents the maximum effect of a single BEF on the outcome.  

\par Finally, we define the spatial scale as the distance at which the effect of a BEF becomes negligible.  This distance is thus intrinsically tied to $\mathcal{K}_s$, and can be calculated upon successful estimation of $\theta^s$ by specifying what is meant by negligible. Operationally, negligible can be defined as a proportion  $p \in (0,1)$ of the maximum effect, which occurs at distance 0 by definition of $\mathcal{K}_s$. The estimated spatial scale $d^{*}$ is then $\hat{d}^{*}=\mathcal{K}^{-1}(p,\hat{\theta}^s) $. 

\subsection{Repeated Measures Model} We now extend this framework to model temporal exposure in a setting where the $i$th subject has $n_i$ visits at which the outcome is measured. Consequently we model $\bm{\mu}_i \in \mathbb{R}^{n_i}$  as a function of the previously mentioned intercept and covariates $\bm{Z}_i$, in addition to subject level effects $\bm{b}_i$ and corresponding design matrix $\bm{W}_i$. To incorporate the temporal dimension of exposure to BEFs, the exposure covariate now aggregates over the exposure times for each subject-BEF pair, in addition to the spatial distances:
    \begin{align*}
    g(\mu_{ij}) &=  \alpha +  \beta X_{ij}(\bm{\theta})  +\bm{Z}_i\bm{\delta} + \bm{W}_i\bm{b}_i, \tag{2}\\
    \text{where} & \\
    X_{ij}(\bm{\theta}) &= \sum_{(d,t) \in \mathcal{D}_{ij}}  \mathcal{K}_s \big( d,\theta^{s} \big )\mathcal{K}_t(t,\theta^{t}) \quad j = 1,...,n_i,
    \end{align*}
\noindent is the spatial and temporal aggregated predictor at the time of study visit $j$ and
$\mathcal{D}_{ij}$ is the set of tuple times, $t\in[0,\infty)$ and distances, $d \in[0,\infty)$ between subject $i$ and the BEFs of interest at occasion $j$. The parameters for BEF of interest are $\bm{\theta} = (\theta^s,\theta^t)$ where $\theta^{t}$ is in the same domain as $t$ and governs the rate at which temporal exposure to a given BEF accumulates according to $\mathcal{K}_t$.  In contrast to $\mathcal{K}_s$, the temporal exposure function $\mathcal{K}_t$ is chosen so that $\mathcal{K}_t(0,\theta^t)=0,\lim_{t\to \infty} \mathcal{K}_t(t,\theta^t)=1$, reflecting the assumption that the maximum effect of a BEF occurs once the subject spends an infinite amount of time near the BEF and vice versus. Thus, similar to how any survival function of a positive random variable may be used as $\mathcal{K}_s$, any cumulative distribution function (cdf) of a positive random variable may be used as $\mathcal{K}_t$. Given the definitions of $\mathcal{K}_s(\cdot)$ and $\mathcal{K}_t(\cdot)$, it follows that $\beta$ represents the change in $g(\bm{\mu}_i)$ when a given BEF is placed at distance 0 from the subject, for an amount of time that approaches infinity.  
\par It is worth noting that, depending on the value of $\theta^t$, $\mathcal{K}_t(t,\theta^t)$ will effectively evaluate to 1 sooner than in the limit to infinity. Denoting $t^{*}$ as the temporal counterpart to the spatial scale discussed earlier, $t^{*}$ is defined as the time at which the exposure function is $\mathcal{K}_t(t^{*},\theta^t)=1-p$, for a small precision $p$. The value $t^{*}$ is interpreted as the time at which a single BEF reaches its highest impact. It can be solved for in a similar manner, setting $ t^* = K_t^{-1}(1-p,\hat{\theta}^t)$. 
\par Any application of this model to varying exposure over the differing visits $k$ will implicitly assume that the coefficient $\beta$ of STAP exposure is equivalent within and between subjects. A new model formulation is required to explicitly separate these effects, a matter which we discuss next.
    
\subsection{Difference in Differences Formulation}
In analysis of repeated measures data, it is often desirable to estimate both the within and between subject effect associated with time-varying covariates. Decomposing these effects can avoid introducing bias into the model coefficients when the effect of within person-changes in exposure are not equal to the association of between-person difference in the exposure and the outcome. This decomposition applies to both the exposure of interest as well as adjustment for time varying covariates. Additionally, the decomposition can alleviate bias concerns due to unmeasured time invariant confounders, and can enable the interpretation of the within subject association as a  causal effect \citep{neuhaus1998between,morgan2013handbook}. For example, in the built environment literature, one confounder that is seldom measured is individual's residential preference, which evolves slowly over time, if at all, and could be considered constant during the study period. While between-person differences in built environments could be counfounded by unmeasured residential preferences, i.e., due to residential selection bias, the within-person change in built environment characteristics is free of this bias. Consequently, this model specification has become increasingly popular in the built environment literature \citep{hirsch2014change}. 
\par To estimate within subject effects, $\beta^w$, STAP exposure components can be centered by the corresponding subject's mean exposure, $\bar{X}_{i}(\bm{\theta})$, to create the new covariate, $\Delta X_{ij}(\bm{\theta})$, which reflects the deviation from average exposure at occasion $j$. Estimation of the between subject effect, $\beta^b$ can be accomplished by including the latent subject specific mean exposure:

\begin{align*}
    g(\mu_{ij}) &= \alpha + \beta^b \bar{X}_{i}(\bm{\theta}) + \beta^w\Delta \bm{X}_{ij}(\bm{\theta})  +    \bm{Z}_i\bm{\delta} + \bm{W}_i\bm{b}_i\tag{3} \label{Eq:DnD} \\
    \bar{X}_{i}(\bm{\theta}) &= \frac{1}{n_i} \sum_{j=1}^{n_i}\sum_{(d,t) \in \mathcal{D}_{ij}} \mathcal{K}_s \big( \frac{d}{\theta^{s}} \big )\mathcal{K}_t(\frac{t}{\theta^{t}}).\\
    \Delta X_{ij}(\bm{\theta})  &= \sum_{(d,t) \in \mathcal{D}_{ij}} \mathcal{K}_s \big( \frac{d}{\theta^{s}} \big )\mathcal{K}_t(\frac{t}{\theta^{t}}) - \bar{X}_{i}(\bm{\theta})\\
\end{align*}

Although a standard GLMM would be able to fit a model with a difference in differences parameterization using a simple pre-processing step, STAP models must be fit with the calculation of the centered covariates built-in to the overall model estimation since the constructed exposures are themselves estimated.

\section{Estimation, Prior Choices and  Model Selection}

We fit the model formulated above in a Bayesian paradigm using MCMC to draw samples from the posterior distribution. We use the software in our \texttt{R} package \texttt{rstap} \citep{peterson2018rstap} which uses the probabalistic programming language \texttt{stan} \citep{StanManual} to implement the Hamiltonian Monte Carlo (HMC) variant ``No U-Turn'' Sampler (NUTS) \citep{hoffman2014no}. A gradient based method such as NUTS is crucial for our modeling framework, as more traditional methods like Gibbs or Metropolis-Hastings Samplers are either not feasible or inefficient.
\par Existing literature on prior choice for standard regression coefficients and auxiliary variables applies here for all STAP parameters other than the spatial and temporal parameters, $\theta^s,\theta^t$ \citep{gelman2007data,gelman2013bayesian}. Priors for spatial and temporal parameters need to be set on the basis of context-specific knowledge, including an understanding of the distribution of distances and times that will be included in the model. This is critical for ensuring model convergence as the model may be poorly identified if one were to use, for example, an improper prior that places equivalent probability mass on all spatial scales, $d^{*}$. Indeed, while probability theory guarantees a valid posterior distribution as long as proper prior distributions are used, this does not guarantee identifiability in a broader sense for our modeling approach. For example, if $\beta$ in (1) above is 0, then the posterior distribution of $\theta^s$ will be equivalent to its prior distribution. Consequently, as is true of Bayesian models more generally \citep{gelman2017prior,san2010bayesian}, our proposed models require forethought to be given as to how priors for the spatial and temporal scales are selected in the context of a given data set or study location so as to arrive at a sensible posterior distribution.  To further illustrate, consider a study area that could be encapsulated within a circle of 10km in diameter, with the \textit{a priori} assumption being that the maximum plausible spatial scale that could be estimated from the model would be 10km, as distances beyond that are not plausible in that study area.
Thus the prior for the spatial parameter $\theta_s$ would be defined so that plausible spatial scales, $d^*=\mathcal{K}^{-1}(p,\theta_s)$, are at most 10km. On the other hand, if the BEF is a destination that a typical person would visit exclusively by walking, then the prior for the spatial parameter should be selected so that the corresponding spatial scales are centered at a distance that is walkable by the average person in the study sample (e.g. 1km).  On the other hand, if the study area is one where the dominant mode of transportation is by car, then the spatial parameter should be selected so that the corresponding spatial scale is larger but remains bounded by the maximum distance possible within the study area. We illustrate this explicitly in our analysis of MESA data.
\par Given that model estimation occurs under a Bayesian paradigm via MCMC, there are a vast array of model validation and selection techniques that can be performed using, for example, posterior predictive checks \citep{gelman2013bayesian} or the Widely Applicable Information Criterion (WAIC) \citep{vehtari2017practical}. The latter is demonstrated in Section 6.

\section{Simulations}
In the present section we demonstrate the STAP model's ability to recover parameter estimates under two broad simulation scenarios:  (Section 5.1) differing spatial patterning of BEFs and (Section 5.2) differing spatial exposure functions. In the latter, we compare our model to the DLM\citep{baek2016distributed}. In both cases we simulate locations of subjects and BEFs within a 2 by 2 square, and outcome data under the following linear model that generates a pseudo BMI for ease of interpretation and analysis:

\begin{align}
    BMI_i &= 23 -2.2 Z_{i} + .75X_{i}(\theta^s=.5) + \epsilon_{i}, \tag{4}\\
    \epsilon & \sim N(0,\sigma^2=1), \nonumber
\end{align}

\noindent where $X_i(\theta)$ is as specified in equation (1). In section 5.1 we use the Exponential exposure function to construct $X_i(\theta^s)$, and vary the spatial distribution of BEFs, thereby create different distributions of distances, $d$.  For the simulations in Section 5.2), we vary the spatial exposure functions, $\mathcal{K}_s$. While we could use any combination of spatial, temporal, or spatial-temporal predictors for the purposes of demonstrating the properties of this modeling paradigm, we primarily examine spatial aggregated predictors for brevity since we found the results extend to the more complicated cases when constructing our \texttt{rstap} package\citep{peterson2018rstap}.  
\par For each simulation replicate, the following model is estimated:

\begin{align*}
    Y_i &= \alpha + \delta Z_{i} + \beta X_{i}(\theta^s) + \epsilon_{i}\\
    \epsilon_i & \stackrel{iid}{\sim} N(0, \sigma^2)\\
    \alpha &\sim N(25,4) \quad \quad \delta  \sim N(0,3) \\
    \beta &\sim N(0,3) \quad \quad \sigma \sim C^+(0,5)\\ 
    \theta^s \sim  &\text{ Gamma}(8,16) \text{ or }\log(\theta^s) \sim N(0,1)
    \\
\end{align*}

\noindent Priors for covariate coefficients, $\bm{\delta}$ and $\bm{\beta}$ are set from recommendations on weakly informative priors \citep{gelman2013bayesian}. 
The intercept prior is set so that the true value, 23, is within three standard deviations of the prior mean, 25.
We use two different priors for scale parameter $\theta^s$ to illustrate the impact of how the prior can affect inference. When using an exponential exposure function $\mathcal{K}(\cdot)$, using precision parameter $p=0.01$, the $\text{Gamma}(8,16)$ prior is equivalent to a prior for $d^*=\mathcal{K}^-1(p, \theta^s)$ that is centered around $\sim 2$; while its support is $\mathbb{R}^+$, it has $95\%$ of its mass between 1 to 4 with the majority of it ranging from $\approx$ 0.2 to $\approx$ 9.2. The more familiar log normal prior yields an overall flatter prior for $d^*$, encompassing both smaller and larger values of the spatial scale compared to the Gamma prior, and is thus deemed as a more conservative prior (see Supplementary Figure S4). 
This demonstrates how, in the absence of other prior information, long tailed priors can be used to estimate $\theta^s$ with good results if the effect, $\beta$, can be detected. The analysis of the MESA data demonstrates the use of priors with the majority of their mass within the range of observed distances as discussed in Section 4.

For each simulation replicate, we use our R package \texttt{rstap}(v1.1.7)  on a Linux Centos 7 operating system with 2x3.0 GHz Intel Xeon Gold 6154 processors, after 2000 burn in iterations we draw 2000 posterior samples  across 4 independent chains. The relatively smaller number of burn-in iterations is a feature of HMC samplers, which converge with fewer iterations.

\subsection{Spatial Patterning}
 To illustrate the robustness of STAP to differing spatial patterns in the locations of subjects and BEFs, we simulate distances, $d$, from settings where the locations of businesses and subjects are either independent or correlated with one another. The correlated scenario reflects the idea that people and BEFs may be collocated near city centers, for example, as would likely be observed in real data.  Specifically, locations are simulated via two-dimensional (1) Homogeneous Poisson process (HPP) and (2) Non-Homogeneous Poisson Process (NHPP).  Since the number of locations can be explicitly specified under the HPP but not the NHPP, the mean of the latter was matched to the fixed number of the former. Plots showing typical simulated spatial patterns from these processes can be seen in Figure \ref{fig:simIeda}. Their corresponding distance distributions can be seen in Figure 2 in the Supplementary Material. In all cases, the exposure function is the Exponential distribution's survival function  (see Figure \ref{fig:simIexp}).
\par Nonzero credible interval coverage, absolute difference, and model calibration are assessed by computing coverage, the percent difference between true and median parameter estimates, interval length, and Cook \& Gelman Statistic \citep{cook2006validation}. We chose these measures since we believe precise inference, as opposed to prediction, is likely of greatest interest to investigators in this field.  We provide a reference to reproducible code in the Supplementary section.

\begin{figure}[H]
    \centering
    \includegraphics[width = .495 \textwidth]{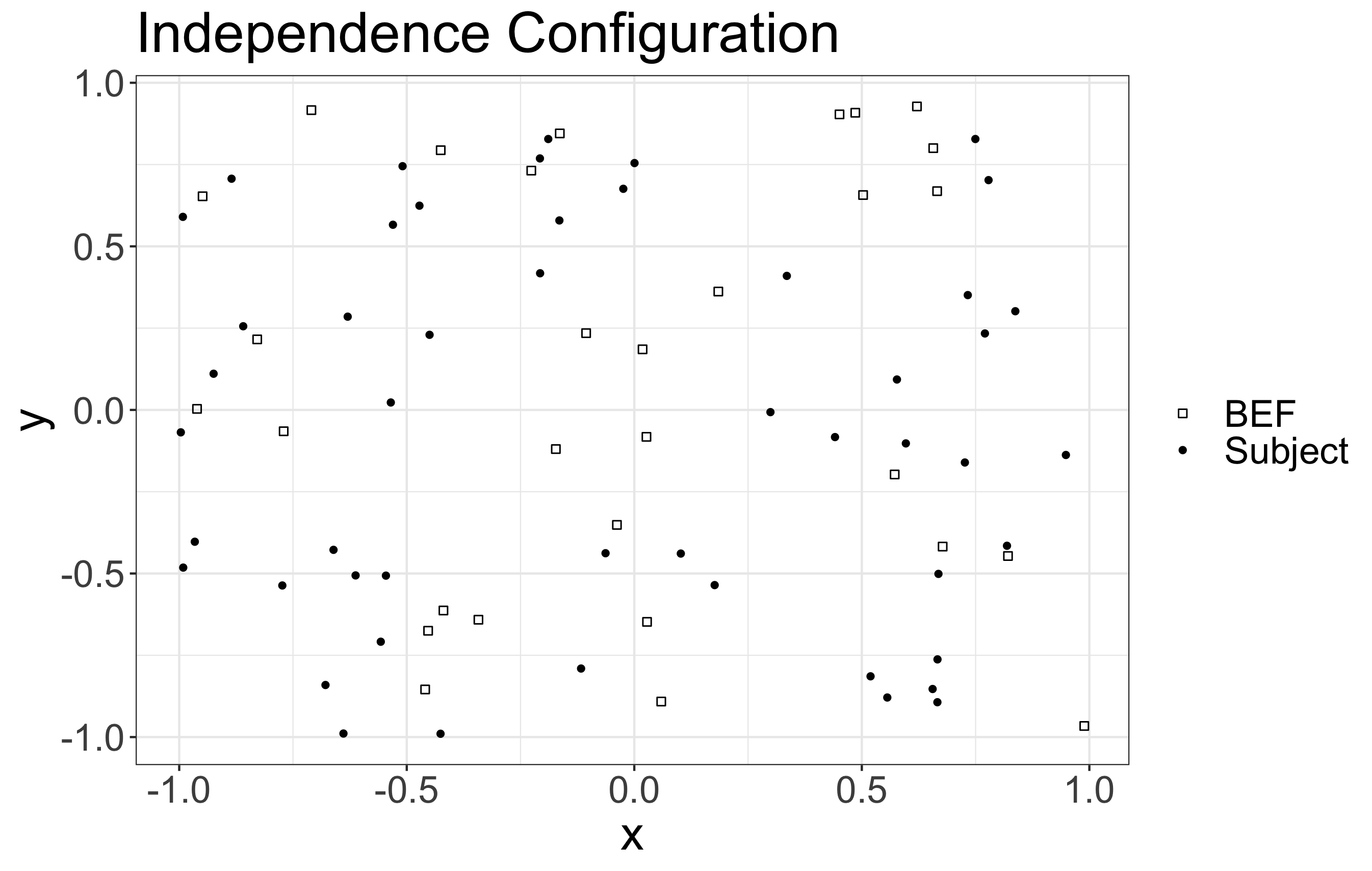}
    \includegraphics[width = .495 \textwidth]{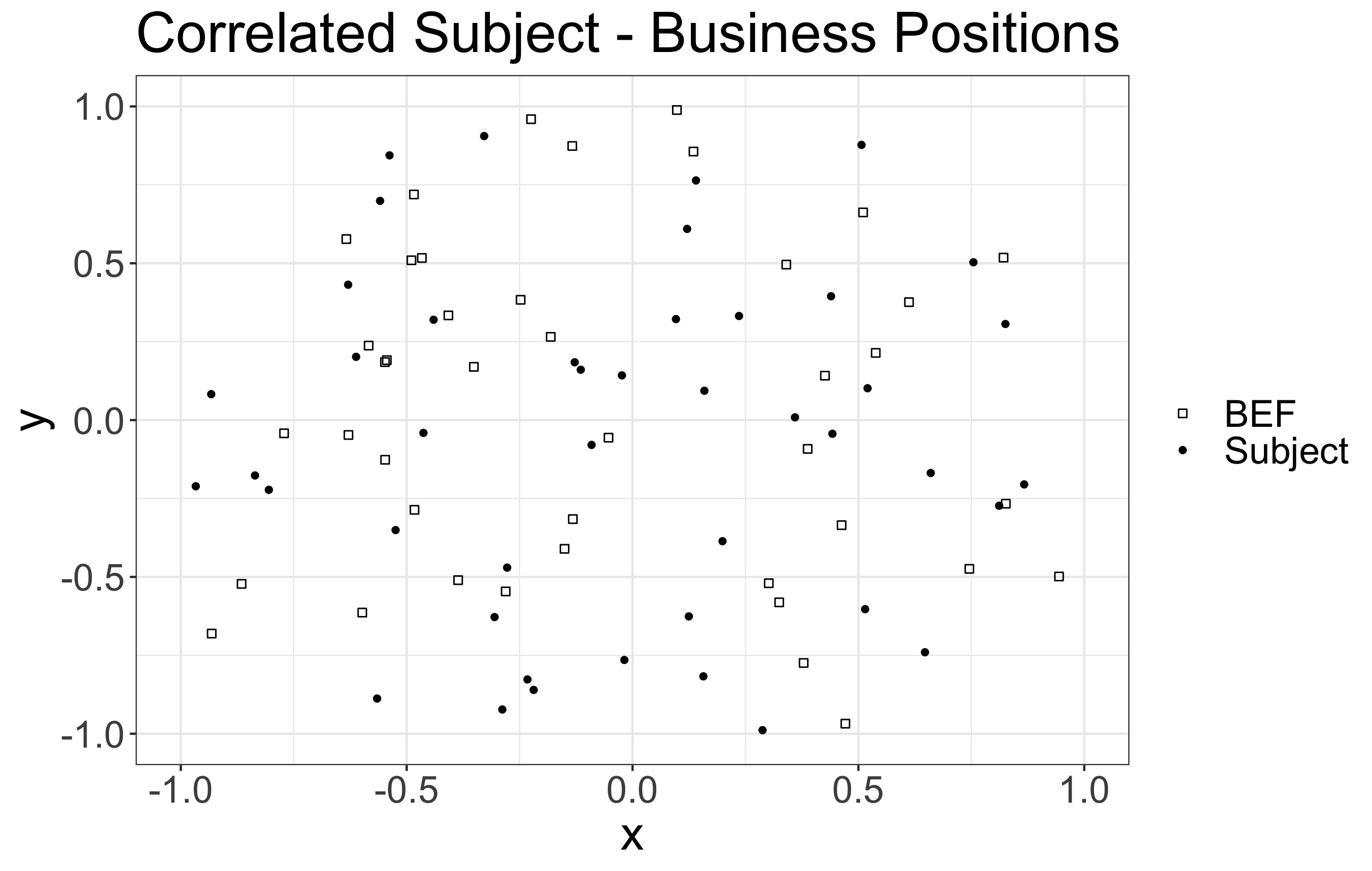}
    \caption{Differing Spatial Arrangement of Subjects and BEFs.}
    \label{fig:simIeda} 
\end{figure}

\subsection{Spatial Exposure Functions}

\begin{figure}[H]
    \centering
    \includegraphics[width = \textwidth]{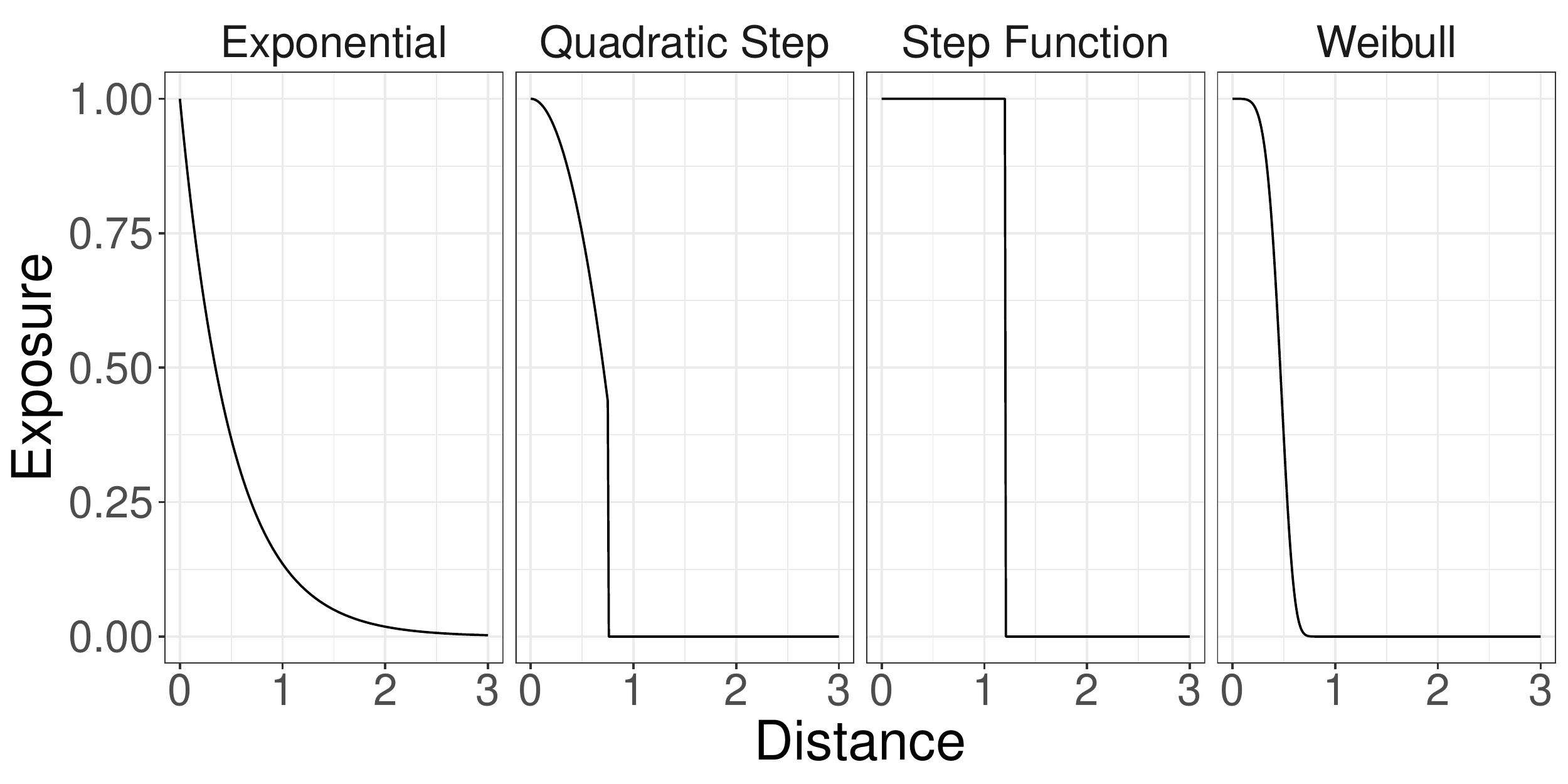}
    \caption{Spatial Exposure Functions. The spatial scales for each exposure function at precision = 0.01 are (left to right) 2.3, .75, 1.2, and 0.68 distance units.}
    \label{fig:simIexp}
    \end{figure}

In order to test the explicit modeling assumption made in regard to the spatial exposure function, we simulate data using the same model as in (5) under the HPP distance distribution using each of four different exposure functions (Figure \ref{fig:simIexp}).

    These exposure functions are chosen to show robustness of STAPs to deviations from assumed exposure functions. Both the Exponential and Weibull distributions survival functions can be modeled explicitly as the STAP spatial exposure function, while step functions of any kind cannot because of their problematic derivatives. Similarly, we combine a quadratic decay function with a step function to create a challenge for the Exponential and Weibull having to model both the decay of the exposure effect over distance as well as a sudden end in the exposure effect.
    \par
    Generating $r=1,...,50$ data sets under each of these spatial exposure functions, STAP models are fit to the data using both the Exponential and Weibull spatial exposure function, using the same priors as detailed at the start of Section 5. From each model fit, we estimate the spatial scale, $\hat{d}^{*}$, using precision $p=0.05$ for the STAP models. The DLM approach of \cite{baek2016distributed} is used as a comparison. Briefly, this approach consists of estimating a set of smoothed coefficients $\gamma(d_\ell)$, for the model $E[Y_i]=\gamma_0 + \sum_{\ell=1}^{L}\gamma(d_\ell) X(d_{\ell-1}; d_\ell)$ where $d_\ell$, $\ell = 1, ..., L$ is a discrete grid of distances, and $X(d_{\ell-1}; d_\ell)$ is the count of BEFs within a ring-shaped area defined by concentric circles of radii $d_{\ell-1}$ and $d_\ell$.  We used $L=20$, and the maximum distance $d_L=3$. For the DLM, we estimate $\hat{d}^{*}$ as the smallest distance for which the credible interval of $\gamma(d_\ell)$ contains zero.  For STAP models, we also calculate the estimate of the effect of the BEF at distance 0, $\hat{\beta}$ – note that there is no directly comparable counterpart to this parameter in the DLM. For each relevant model fit and generative model combination we calculate the mean percent absolute difference from the true spatial scale and effect, e.g. $\frac{1}{50} \sum_{r=1}^{50}\frac{|\hat{d}_r^*-d^*|}{d^*}$, which are shown in Figure \ref{fig:sims2}.

\begin{figure}
    \centering
    \includegraphics[width =\textwidth]{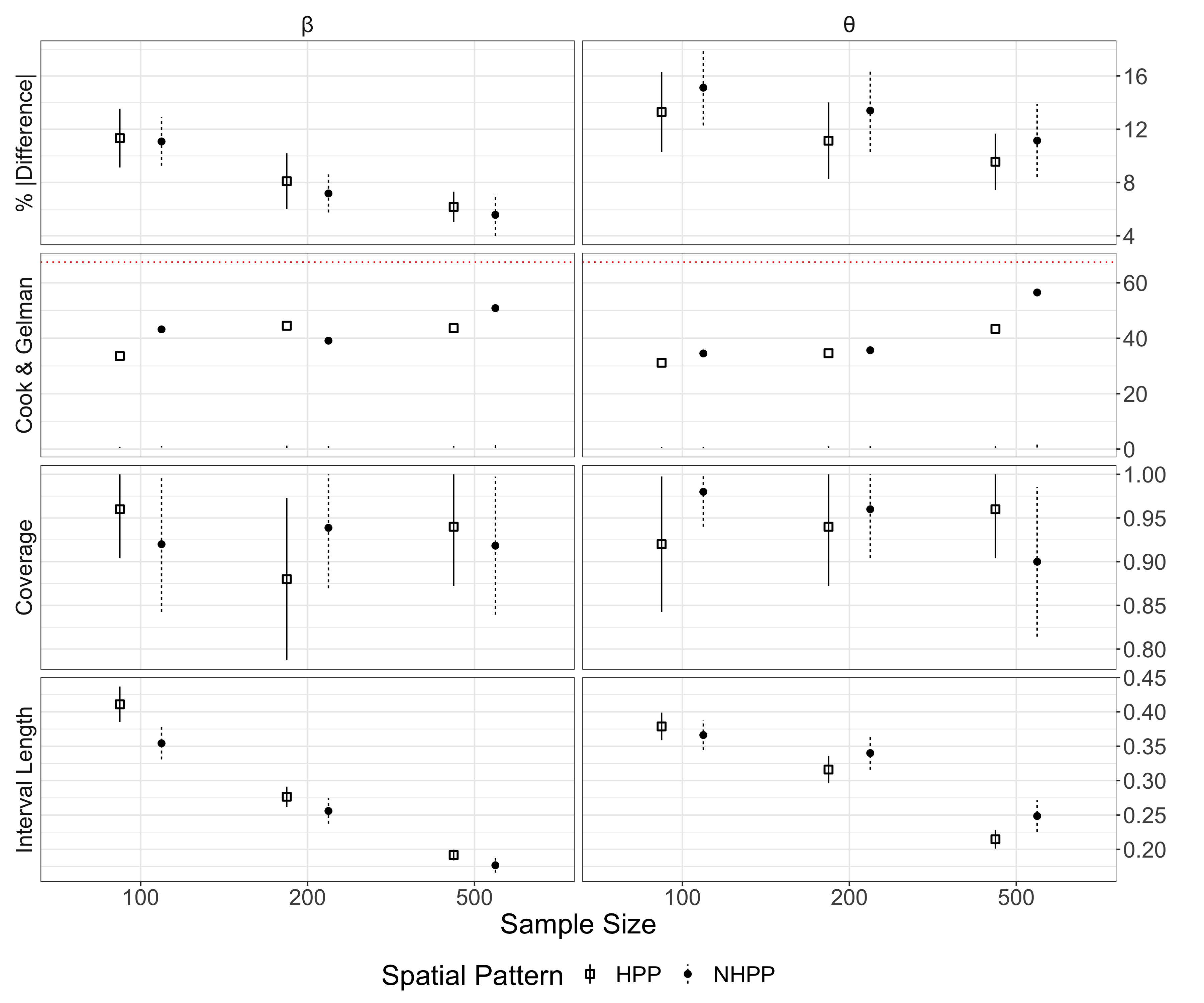}
    \caption{Simulation Results Evaluated by (Top Row) Absolute Difference, (2nd Row) Calibration Statistic (3rd Row) Coverage and (4th) Interval Length across Sample Size and Spatial Pattern, using the $\theta^s\sim$Gamma(8,16) Prior. For all plots but Cook \& Gelman, dots and intervals indicate median, 95\% credible interval respectively. Results for the log$(\theta^s)\sim$Normal(0,1) prior appear in the Supplementary Material.}
    \label{fig:sims1}
\end{figure}

\begin{figure}
    \centering
    \caption{Percent absolute difference in median estimate of (Top) effect size $\beta$ and (Bottom) spatial scale $d^*$ from simulations varying information in sample size, generated spatial exposure function and modeled spatial exposure function. Panel title indicates the generating spatial exposure function, while dot shape indicates the median absolute difference for the modeled spatial exposure function. Line width is the 95\% credible interval.}
    \includegraphics[width =  \textwidth]{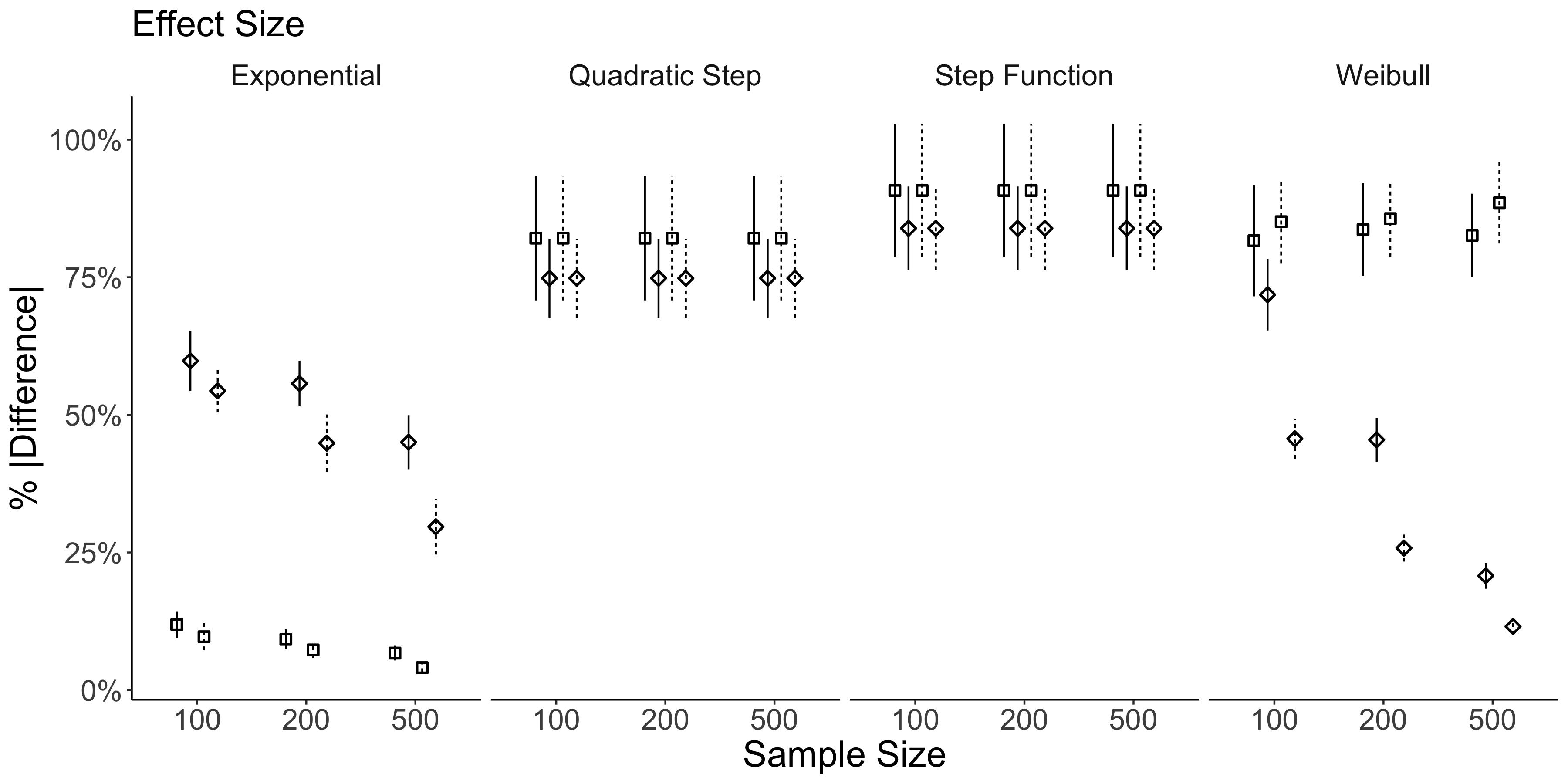}
    \includegraphics[width = \textwidth]{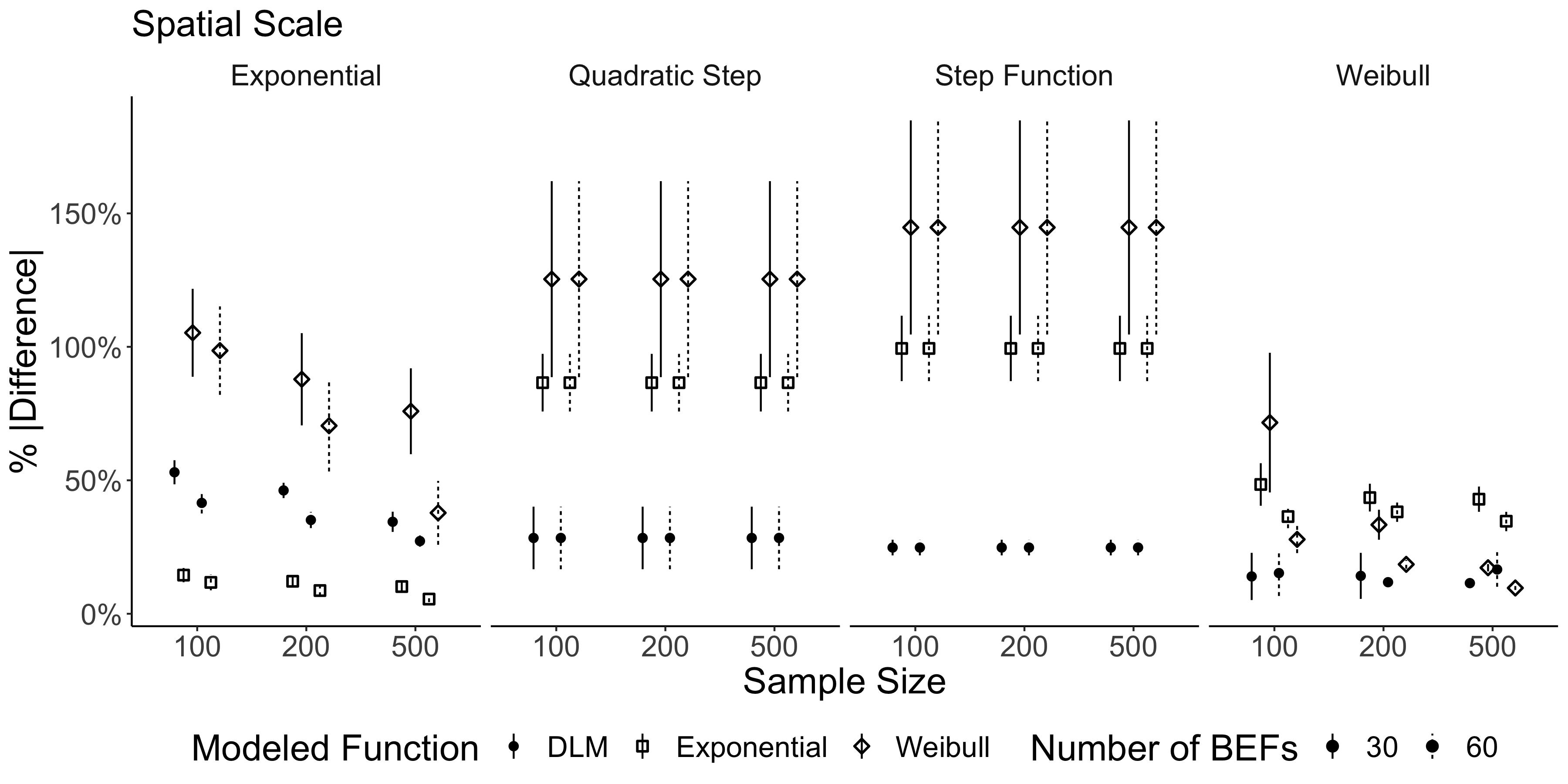}
    \label{fig:sims2}
\end{figure}

\subsection{Simulation Results}

Figures \ref{fig:sims1} and  \ref{fig:sims2} show simulation results when using the $\theta^s\sim$Gamma(8,16) prior. Figure \ref{fig:sims1} demonstrates the properties of STAP model estimators behave as expected in our simulated conditions. That is, as sample size increases, point estimate error and interval length decreases, while empirical coverage rates and model calibration remain constant. In examining the differences in behavior between the two different spatial patterns, we see that the primary difference between them is that correlated subject-BEF distances regularly result in a higher bias in the estimated spatial parameter as compared to the non-correlated subject-BEF distances at a given sample size.  
\par  Turning our attention to Figure \ref{fig:sims2}, we see that the Weibull function performs better than the Exponential in estimating the spatial scale, $d^*$ and exposure effect, $\beta$, in almost all cases except the simulated Exponential. This pattern is likely a consequence of the increased flexibility that comes from the Weibull's shape parameter: in estimating the step functions, this shape parameter provides the STAP model the flexibility to more closely approximate the step function's constant or near-constant exposure effect at shorter distances. However, the Weibull will of course not be able to as efficiently estimate an Exponential exposure function, since the shape parameter would have to be estimated at 1 with low uncertainty in order to achieve similar results as its Exponential counterpart. The results from using the log$(\theta^s)\sim$Normal(0,1) prior are generally similar to those shown in \ref{fig:sims1} and  \ref{fig:sims2}; see Supplementary Material Figures 4 and 5.

\par The simulation results provide guidance for investigators: using the Weibull exposure function is advantageous if there are little to no concerns regarding the ability to detect the exposure effect and there is a low level of substantive certainty regarding the functional form of spatial or temporal exposure. In contrast, if there is a greater need for estimation efficiency and a higher substantive certainty that the spatial exposure function decays exponentially, found via exploratory use of the DLM for example, then the appropriate course of action would be to use the Exponential exposure function. If there remains uncertainty about any of these conditions, model selection tools like WAIC can be used to provide evidence in favor of one exposure function over another, as we demonstrate in Section 6.

\section{Relationship between exposure to healthy foods stores and BMI in MESA}
This section presents the results from fitting the STAP model to data from the MESA cohort participants residing within North Carolina, to examine the relationship between availability of healthy food stores (HFS), defined in Section 2, and body mass index (BMI). As previously mentioned, we focus on participants who originally enrolled in the North Carolina site because, compared to other MESA sites, the relatively less dense urban environment at this site makes so that spatial scales could be larger than the assumed 1.6 km (1 mile) spatial scale used in prior literature. The objective is to (i)  estimate the effects of HFS availability on average BMI, (ii) estimate the spatial and temporal scales at which this effect occurs and (iii) determine which exposure function better explains the spatial exposure to Healthy Food Stores (HFS). We focus on individuals who moved residences at some point during follow-up, since these individuals are more likely to experience within-person change in BEF exposure and thus may provide information about the association of within-person change in BEF and within person change in BMI (i.e., $\beta^w$ in Equation \ref{Eq:DnD}). 
\par We fit the model shown below, adjusting for standard confounders including age, education, sex, and others (Supplementary Table 2):
\begin{align*}
E[BMI_{ij}|\bm{b}_i] &= \alpha + \beta^w\Delta X_{supermarket,ij}(\theta^s,\theta^t) + \beta^{b}\bar{X}_{supermarket,ij}(\theta^s,\theta^t) \\
&+  \bm{Z}_{ij}^{T}\bm{\delta} \\ 
&+b_{i1} + b_{i2}t_{ij}.
\end{align*}

\noindent We fit two models in R \citep{Rlanguage} using both the Weibull and Exponential Spatial Exposure functions, via the \texttt{rstap}(v1.1.06) package (v.4.0.2) on a Windows 10 operating system with a 3.7 GhZ Core i9 Intel Processor. We run four independent chains drawing 2000 samples from the posterior distribution after warming up the sampler for 2000 samples on each chain. Priors on regression coefficients were normal with mean 0 and standard deviation 3, which were selected following standard recommendations  \cite{gelman2013bayesian}. For the spatial parameter, $\theta^s$, we use a Log Normal prior with its scale set at 0.3, to reflect the evidence from previous work \citep{baek2016hierarchical} that probable spatial scales are lower than 5 km. Specifically, when using the Exponential survival function to model $\mathcal{K}$, the median of the corresponding prior the spatial scale, $d^{*}_{.5}$ is at $\approx$ 4.6 km for proportion of effect $p=0.01$. This spatial scale is in line with the estimates in the cited work. For both models we use the exponential cdf to estimate temporal exposure using the same Log Normal$(0,.3)$ prior for the temporal parameter, $\theta^t$. As there is little previous research in estimating temporal scales, this prior is selected based on the  fact that the prior encodes that belief  that, with 50\% probability,  the majority of the effect of living near a grocery store occurs within 4.5 years of living close to the store. 
\par
\begin{figure}[H]
    \centering
    \includegraphics[width =   .9\textwidth]{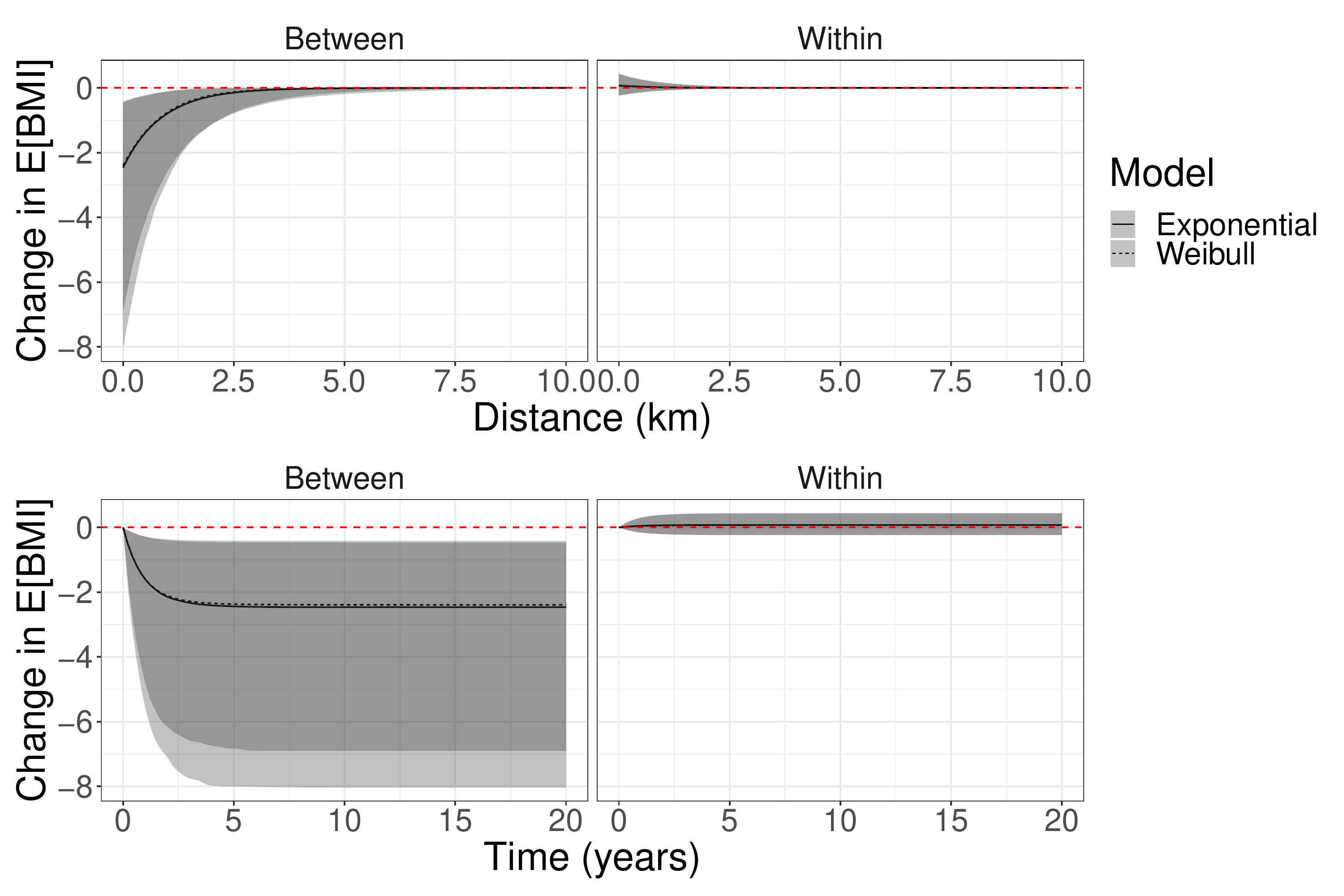}
    \tiny
    \caption{\\ Estimated associations between BMI and healthy food store availability near North Carolina MESA participants' residential locations. (Top) Between- and within-subject associations as a function of distance from models using the Weibull and Exponential spatial exposure functions. (Bottom) Between- and within-subject associations as a function of time, using the Exponential exposure function. Shaded area corresponds to the 95\% Posterior Credible Interval, interior dotted and full lines denote the median estimate of the corresponding spatial exposure function.}
    \label{fig:MESA}
\end{figure}

 We calculate posterior credible intervals for $\beta^w$, $\beta^b$, and the spatial and temporal scales in a standard fashion, namely the median and the 2.5 and 97.5 percentiles of the posterior distributions. In addition, to help visualize the total magnitude of the effect at a given time or distance, we calculate estimates and posterior point wise credible intervals for the product of the regression coefficient ($\beta$) and the respective exposure function. First, for each posterior sample, $m=1,...,M$ of the spatial and temporal parameters ($\theta^{(m)}_s$ and $\theta^{(m)}_t$) we evaluate the models'  spatial-temporal exposure functions along a grid of distance or time values. We then multiply the resulting quantity by the $\beta$ values for the between or within effect from the same posterior iteration $m$. Quantiles are then calculated at each grid value, resulting in a point wise estimate of the total function effect at each point in distance or time. For instance, for distance $d$, the median estimate for the within subject effect across space for the Exponential exposure function would be: $\text{median}_{m}( \beta_m^w \times  \text{exp}(-\frac{d}{\theta_m^s}))$, where the median is taken across $M$  total samples. The resulting functions are  plotted in Figure \ref{fig:MESA}. Note that we use the joint posterior distribution here to fully accommodate uncertainty in all model parameter estimates. However, depending on the question of interest, others may wish to marginalize or condition on the spatial or temporal parameters to arrive at an average or conditional effect estimate.

In the Weibull model, the median between-subject effect, $\beta^b$, was estimated to be -2.39 (95\% CI: -8.04 , -.470) BMI units per unit increase in average exposure, substantially larger than the corresponding  within-subject effect, $\beta^w$,  estimated to be  0.07 (95\% CI: -.23, .44), which is on the scale of deviations from average exposure.   Similarly, the Exponential model estimates were -2.46 (-6.9 ,-.417) and 0.076 (95\% CI: -.23 ,0.43) for the same parameters, respectively. This suggests that a higher average exposure to supermarkets is associated with a lower BMI after adjusting for all relevant confounders. However, there is effectively no evidence for a relationship between change in within-person level supermarket exposure and change in obesity from this model and these data.
\par   Median estimates of the spatial scale, the distance at which the association between HFS exposure and BMI becomes practically negligible, were 4.11 km and 4.2 km in the Exponential and Weibull models, respectively, as found in Table \ref{tab:term}. Both estimates represent a shift slightly leftward, from the median \textit{a priori} spatial scale of 4.6 km. Median estimates for the temporal scale, the time at which the association between HFS exposure and BMI is effectively optimal , were 4.8 years with both Weibull and Exponential models showing a similar rightward shift from the prior temporal scale of 4.6 years.

\par In order to determine which exposure function better describes spatial exposure, we calculate the WAIC for each model fit - see Table \ref{tab:term} -  and find that the Exponential model has a lower WAIC than the model fit with the Weibull spatial exposure function, suggesting the former model has better fit and greater out-of-sample predictive accuracy \citep{vehtari2017practical}.

\begin{table}[H]
\centering
	\begin{tabular}{lll}
  \hline
  & Exponential & Weibull \\ 
  \hline
Spatial (km) & 4.18 (2.42, 7.2) & 4.11 (1.74, 10.39) \\ 
  Temporal (years) & 4.76 (2.67, 8.49) & 4.75 (2.6, 8.37) \\ 
  WAIC & 27,453 & 103,787 \\ 
   \hline
\end{tabular}
	\caption{Estimated Spatial-Temporal  Scales – Median (2.5\%,97.5\%) – at precision $p$=0.01.}
	\label{tab:term}
\end{table}

\section{Discussion}

In this work we motivated, proposed, tested and demonstrated the STAP model family using both simulated data and data from the Multi-Ethnic Study of Atherosclerosis. STAP models are motivated by the need to estimate the spatial and/or temporal scale at which BEFs may effect subjects' health or health behaviors in their environment. While built environment data structures are the primary motivation of this modeling framework, other point pattern spatio-temporal data structures could also be used in this framework. 
\par Similar to the DLM modeling approach \citep{baek2016distributed},  we fully condition on distances constructed from point pattern BEF data in order to estimate a BEF's spatial scale. In contrast to their use of splines to non-parametrically estimate the spatially-varying effect, we assume a constrained functional form of spatial-temporal exposure in order to estimate a latent exposure covariate to use in regression. The use of a constrained functional form is similar to the models developed in \citep{heaton2011spatial,heaton2012kernel}; however, in our approach the exposure surface (point pattern data) is fully observed and does not require modeling.
\par  Our simulations show that little is lost by the use of parametric constraints as the Weibull exposure function is able to estimate the spatial scale of interest with less error than a competing DLM model. However, because there may be insufficient information to support the use of an exposure function as flexible as the Weibull, we've demonstrated how to use tools like WAIC to formally decide between differing exposure functions.
\par There are a number of ways in which STAPs may be extended in order to better answer more complex questions of substantive interest. For example, the spatial-temporal parameters could be constructed to incorporate a more complex hierarchical structure, providing a subject or group specific spatial-temporal scale interpretation to investigators. Additionally, incorporating interaction effects would further strengthen the ability of the STAP family to test epidemiological questions motivating theories of how different BEFs impact groups of subjects differently. For example, the relevant spatial scale may be smaller for older adults with mobility limitations. Finally, employing computational strategies such as consensus Monte Carlo or stochastic gradient descent for these large scale data could decrease the time required to estimate parameters of interest. 
\par 
Although there is a long history of Bayesian spatial-temporal models, this is, to our knowledge, the first adaptation of latent spatial variable models to fully conditioned distance or time data as a covariate in a regression model. Appropriately used, this novel methodology can increase understanding of how the community resources we live around can effect our health.


\section*{Acknowledgments}
This research was supported by NIH grant R01-HL131610 (PI: S\'anchez). MESA research was supported by contracts \\ 75N92020D00001,
HHSN268201500003I, N01-HC-95159, 75N92020D00005, N01-HC-95160, 75N92020D00002, N01-HC-95161, 75N92020D00003, N01-HC-95162, 75N92020D00006, N01-HC-95163, 75N92020D00004, N01-HC-95164, 75N92020D00007, N01-HC-95165, N01-HC-95166, N01-HC-95167, N01-HC-95168 and N01-HC-95169 from the National Heart, Lung, and Blood Institute, and by grants UL1-TR-000040, UL1-TR-001079, and UL1-TR-001420 from the National Center for Advancing Translational Sciences (NCATS).  The authors thank the other investigators, the staff, and the participants of the MESA study for their valuable contributions.  A full list of participating MESA investigators and institutions can be found at http://www.mesa-nhlbi.org.

\section*{Supporting information}

The following supporting information is available as part of the online article:

\noindent
\textbf{Table S1.}
{Sample Data Structure for Subject - Built Environment Data}

\noindent
\textbf{Figure S1.}
{Density Estimate of Healthy Food Store Exposure for North Carolina MESA participants across measurements- indicated by line type- and paneled by buffer sizes. The panel title indicates the buffer size.}

\noindent
\textbf{Figure S2.}
{Simulated Spatial Pattern Distance Distributions}

\noindent
\textbf{Table S2.}
{MESA Subjects descriptive statistics at baseline. $^{1}$Statistics presented: n (\%) ; median (IQR)}

\noindent
\textbf{Figure S3.}
{Priors for spatial scales used in simulations.}

\noindent
\textbf{Figure S4.}
{Conservative prior simulation results Evaluated by (Top Row) Absolute Difference, (2nd Row) Calibration Statistic (3rd Row) Coverage and (4th) Interval Length across Sample Size and Spatial Pattern. For all plots but Cook \& Gelman, dots and intervals indicate median, 95\% credible interval respectively}

\noindent
\textbf{Figure S5.}
{Percent difference in median estimate of (Top) effect size $\beta$ and (Bottom) spatial scale $d^*$ from simulations varying information in sample size, generated spatial exposure function and modeled spatial exposure function using conservative prior.Panel title indicates the generating spatial exposure function, while dot shape indicates the median absolute difference for the modeled spatial exposure function. Line width is the 95\% credible interval.}

\nocite{*}
\bibliography{wileyNJD-AMA}

\clearpage

\end{document}


\maketitle
\centering
\begin{table}[H]
           \centering
            \begin{tabular}{c|c|c|c|c}
                subject\_ID & visit\_ID & BEF    & Distance & Time \\ \hline
                1& 1       & Fast Food   & 0.3 & 3.4    \\
                1& 1       & Fast Food   & 0.8 & 2.8     \\
                1& 1       & Fast Food   & 0.4 & .5    \\
                1& 2       & Fast Food   & 1.3 & .4
            \end{tabular} 
            \caption{ Sample Data Structure for Subject - Built Environment Data }
    \end{table}

\begin{figure}[H]
    \centering
    \includegraphics[width=.8\textwidth]{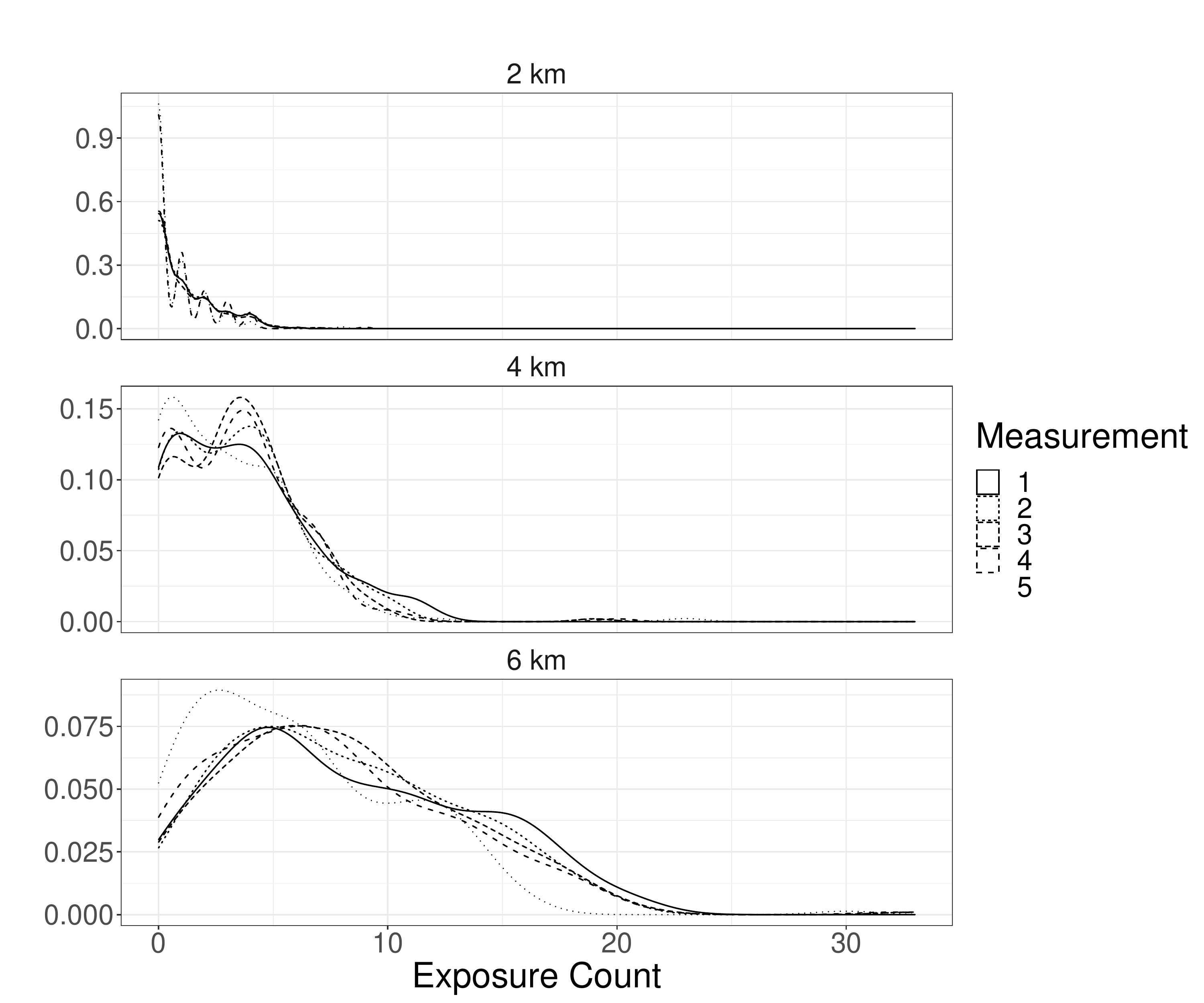}
    \caption{Density Estimate of Healthy Food Store Exposure for North Carolina MESA participants across measurements- indicated by line type- and paneled by buffer sizes. The panel title indicates the buffer size.}
    \label{fig:dhist}
\end{figure}

\begin{figure}[H]
    \centering
    \includegraphics[width =.5\textwidth]{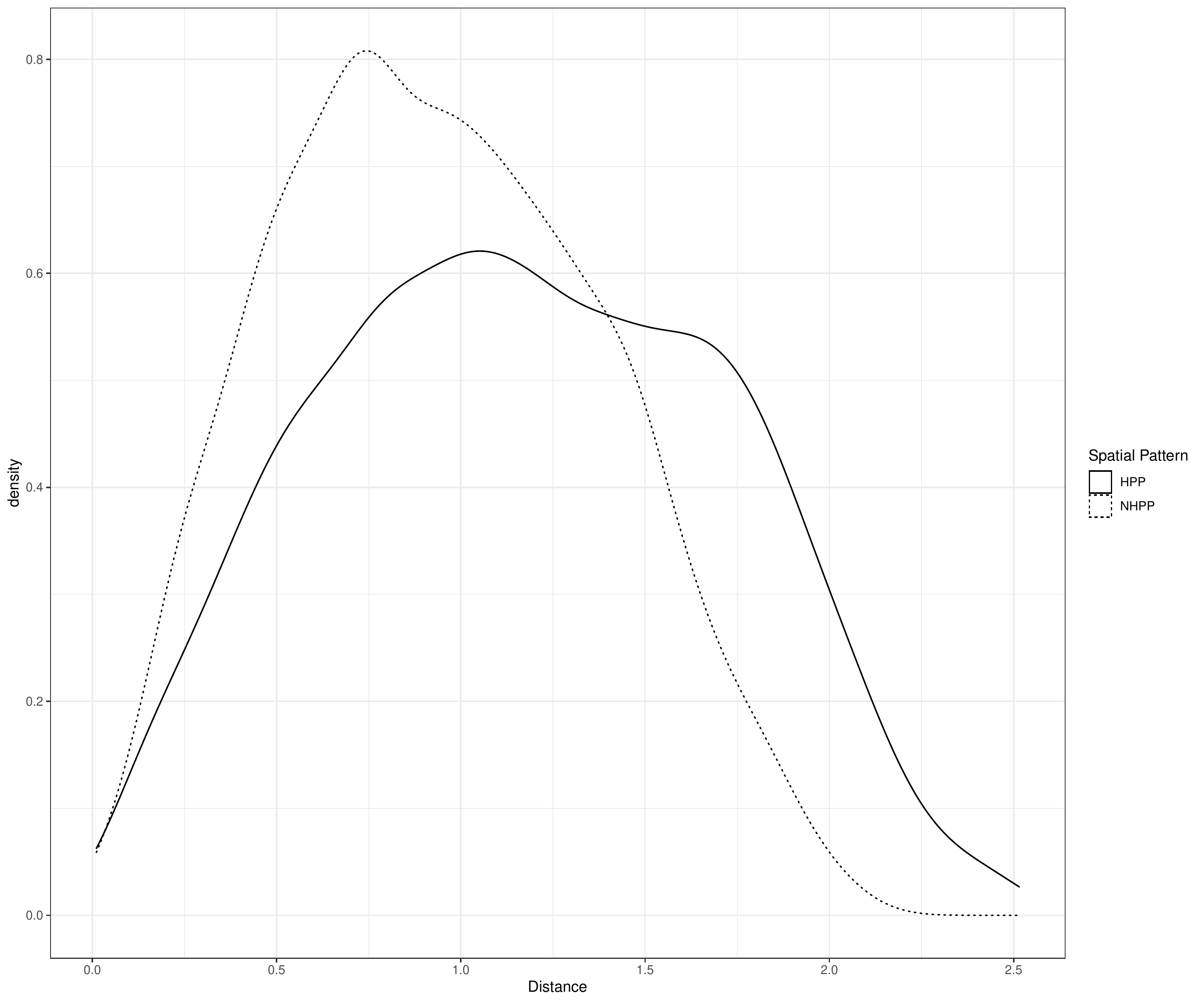}
    \caption{Simulated Spatial Pattern Distance Distributions}
    \label{fig:sim_ddist}
\end{figure}

\begin{figure}
    \centering
    \includegraphics[width = \textwidth]{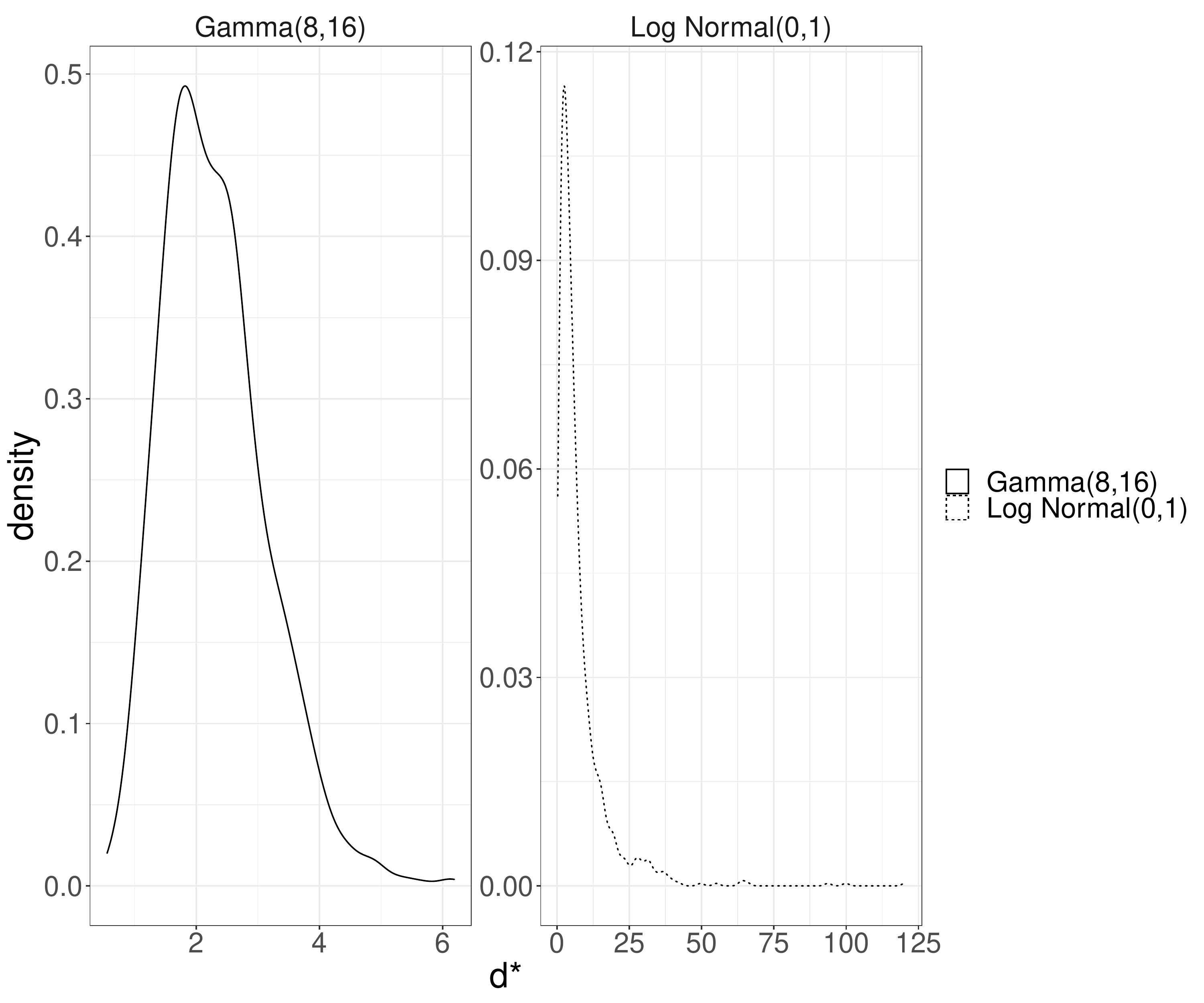}
    \caption{Priors for spatial scales used in simulations.}
    \label{fig:sim_priors}
\end{figure}

\begin{table}[H]
\centering
\begin{tabular}{l|l}
\hline
\textbf{Characteristic} & \textbf{N = 311}$^{1}$  \\ 
\hline
Sex & \\
\hline
\quad Female & 170 (55\%)\\
\hline
\quad Male & 141 (45\%)\\
\hline
Race & \\
\hline
\quad Black & 129 (41\%)\\
\hline
\quad Hispanic & 2 (0.6\%)\\
\hline
\quad White & 180 (58\%)\\
\hline
\# Of Follow Up & \\
\hline
\quad 1 & 2 (0.6\%)\\
\hline
\quad 2 & 16 (5.1\%)\\
\hline
\quad 3 & 59 (19\%)\\
\hline
\quad 4 & 234 (75\%)\\
\hline
Education & \\
\hline
\quad High School/GED (or less) & 75 (24\%)\\
\hline
\quad Some College & 102 (33\%)\\
\hline
\quad Bachelors & 134 (43\%)\\
\hline
Income (1000s USD) & \\
\hline
\quad $<$12 & 17 (5.5\%)\\
\hline
\quad 12-24 & 56 (18\%)\\
\hline
\quad 25-39 & 51 (16\%)\\
\hline
\quad 40-74 & 101 (32\%)\\
\hline
\quad $>$75 & 86 (28\%)\\
\hline
Married & 195 (63\%)\\
\hline
History of Cancer  & 34 (11\%) \\
\hline 
Smoking Status & \\
\hline
\quad Never & 131 (42\%)\\
\hline
\quad Former & 127 (41\%)\\
\hline
\quad Current & 53 (17\%)\\
\hline
Age at Exam & 60 (53, 68)\\
\hline
Body Mass Index (kg/m$^2$) & 28.0 (25.0, 31.6)\\
\hline 
\end{tabular}
\caption{MESA Subjects descriptive statistics at baseline. $^{1}$Statistics presented: n (\%) ; median (IQR).}
\end{table}
\pagebreak
Reproducible code for the simulations can be found at \href{https://github.com/apeterson91/STAPSimulations}{https://github.com/apeterson91/STAPSimulations}.

\begin{figure}
    \centering
    \caption{Conservative prior simulation results Evaluated by (Top Row) Absolute Difference, (2nd Row) Calibration Statistic (3rd Row) Coverage and (4th) Interval Length across Sample Size and Spatial Pattern. For all plots but Cook \& Gelman, dots and intervals indicate median, 95\% credible interval respectively.}
    \includegraphics[width = .9\textwidth]{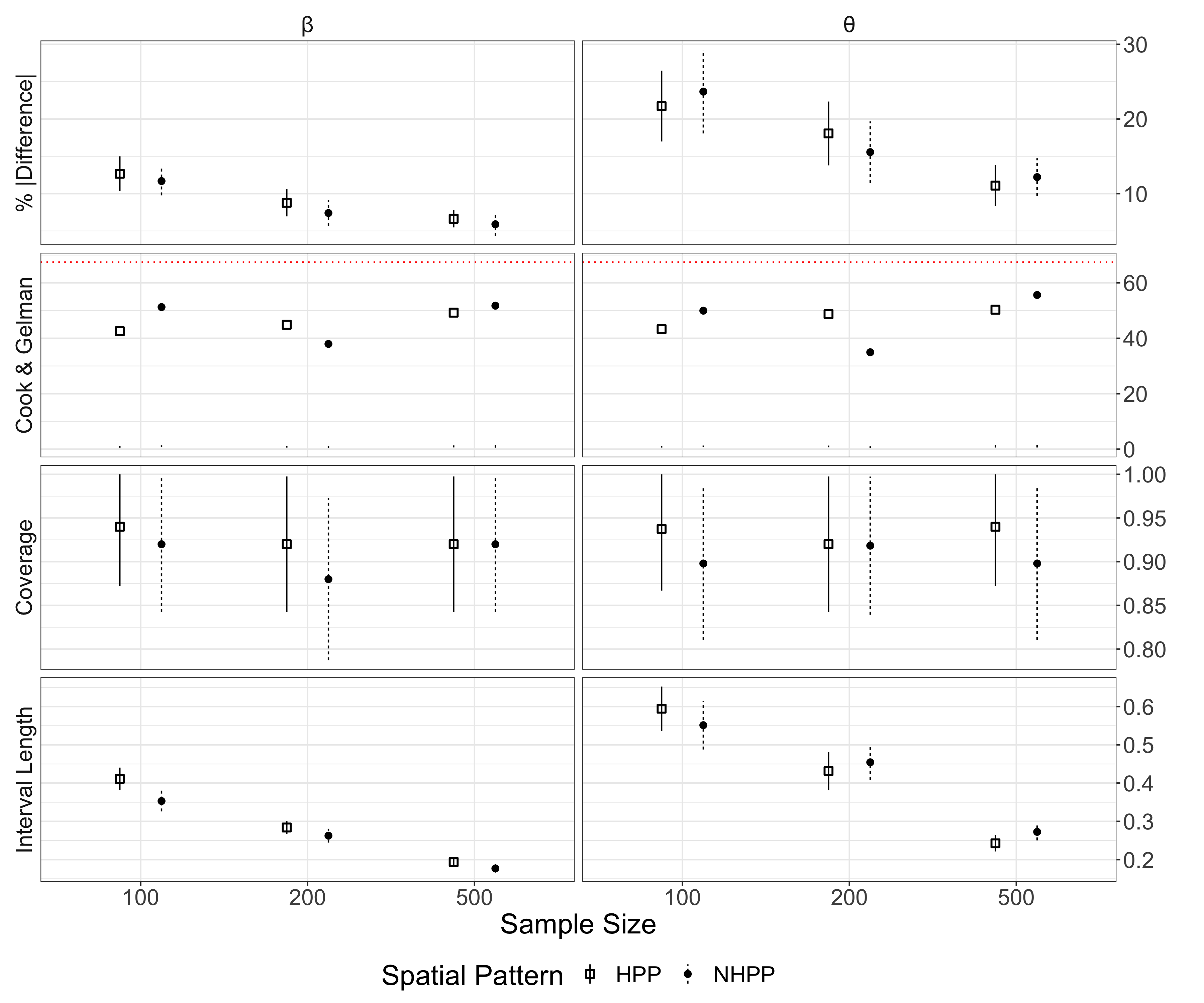}
    \label{fig:lnsim1fig}
\end{figure}

\begin{figure}
    \centering
    \caption{Percent difference in median estimate of (Top) effect size $\beta$ and (Bottom) spatial scale $d^{\star}$ from simulations varying information in sample size, generated spatial exposure function and modeled spatial exposure function using conservative prior.Panel title indicates the generating spatial exposure function, while dot shape indicates the median absolute difference for the modeled spatial exposure function. Line width is the 95\% credible interval.}
    \includegraphics[width =  \textwidth]{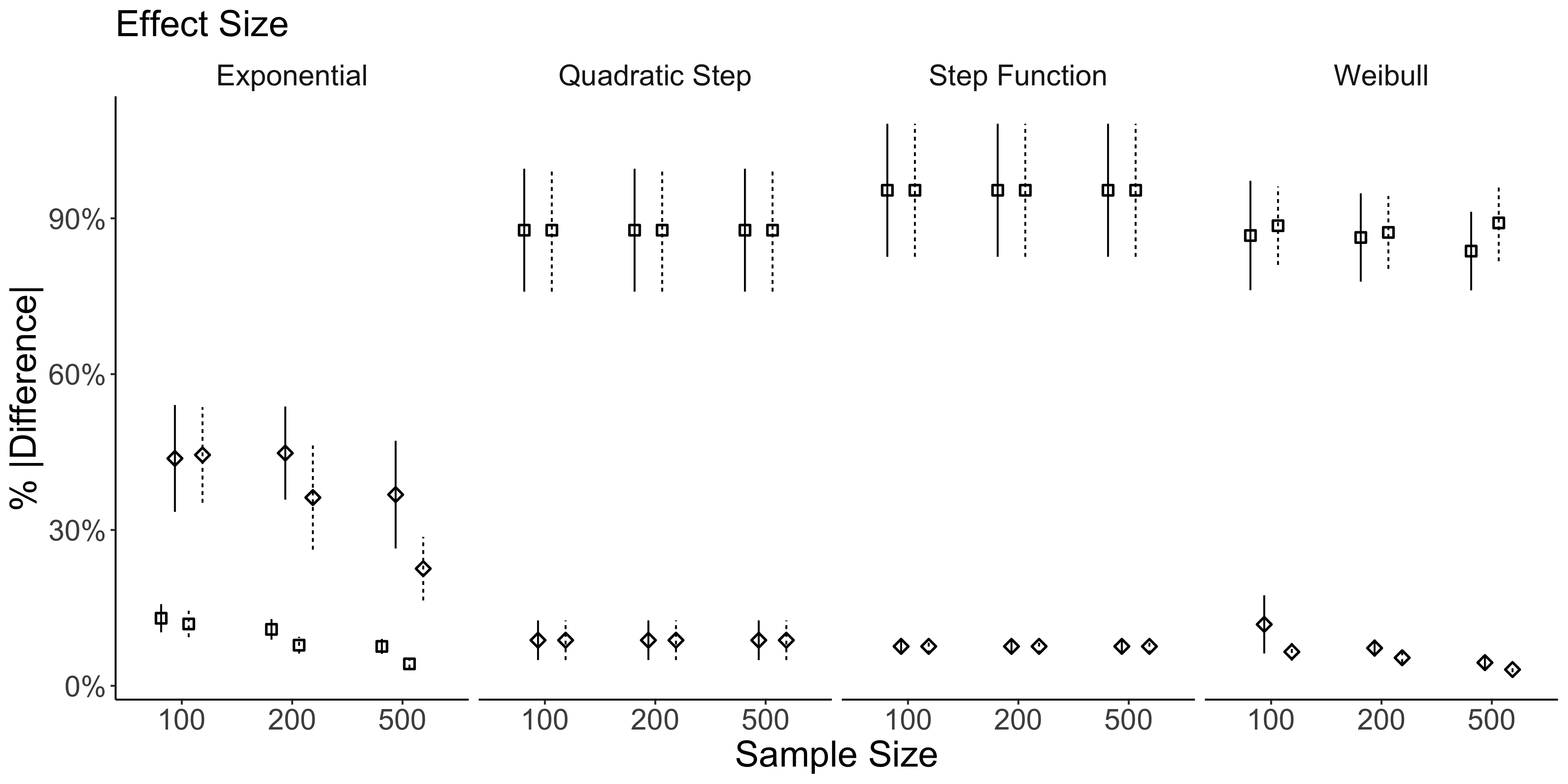}
    \includegraphics[width = \textwidth]{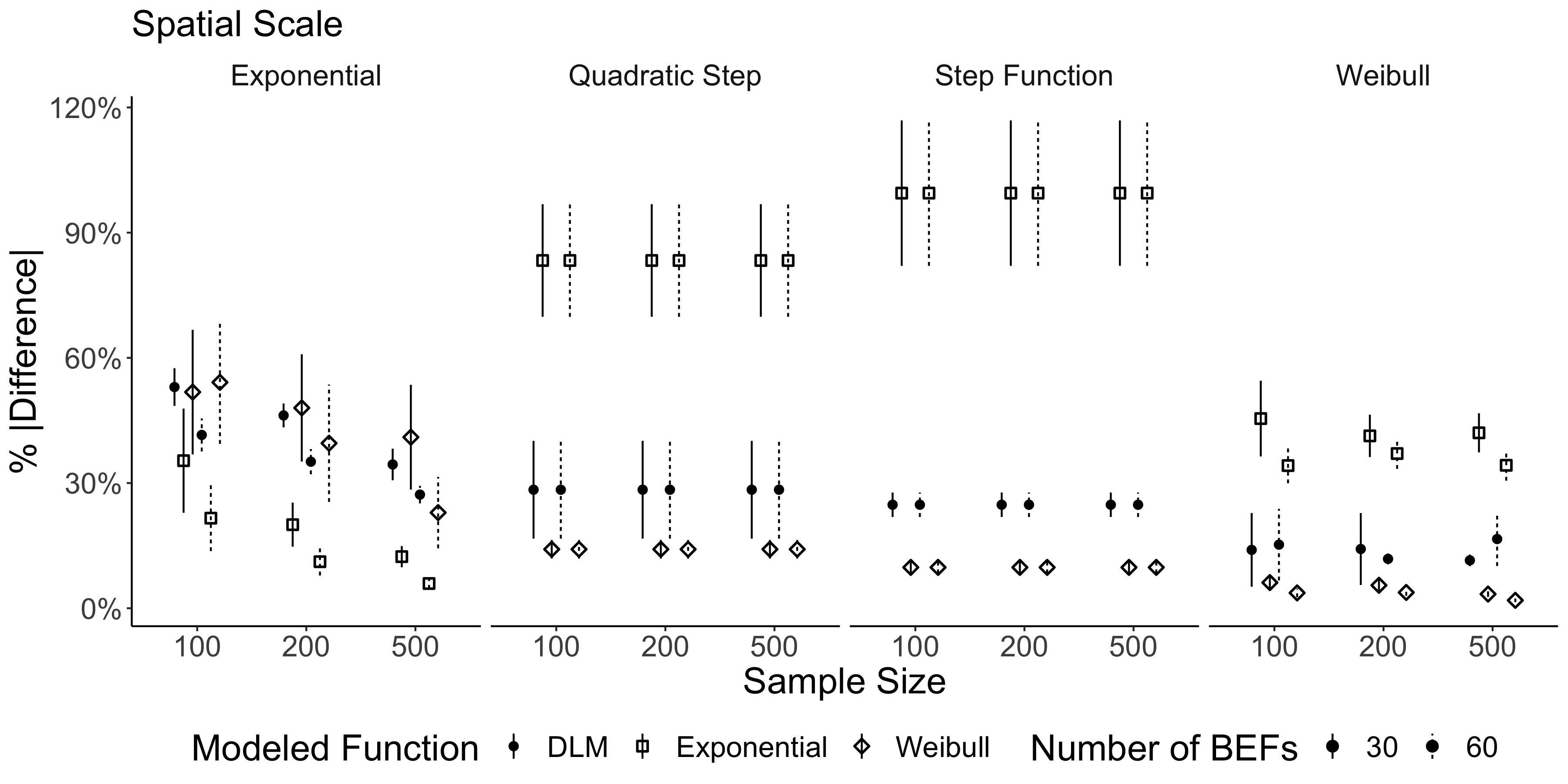}
    \label{fig:lnsims2}
\end{figure}